\documentclass{emulateapj}

\shorttitle{The Origin of [OII] Emission in Recently Quenched AGN Hosts}
\shortauthors{Kocevski et al.}

\begin{document}

\title{The Origin of [OII] Emission in Recently Quenched AGN Hosts}

\author{Dale D. Kocevski, Brian C. Lemaux\altaffilmark{1}, Lori
  M. Lubin\altaffilmark{1}, Alice E. Shapley\altaffilmark{2}, Roy
  R. Gal\altaffilmark{3}, \& Gordon K. Squires\altaffilmark{4}} 

\affil{University of California Observatories/Lick Observatory, University of California, Santa Cruz, CA 95064}
\altaffiltext{1}{Department of Physics, University of California, Davis, 1 Shields Avenue,  Davis, CA 95616}
\altaffiltext{2}{Department of Physics and Astronomy, 430 Portola Plaza, University of California, Los Angeles, CA 90095}
\altaffiltext{3}{Institute for Astronomy, University of Hawaii, 2680 Woodlawn Dr., Honolulu, HI 96822}
\altaffiltext{4}{\emph{Spitzer} Science Center, M/S 220-6, California 
                 Institute of Technology, 1200 East California Blvd, Pasadena, CA 91125}
\email{kocevski@ucolick.org}

\begin{abstract}
We have employed emission-line diagnostics derived from DEIMOS and NIRSPEC
spectroscopy to determine the origin of the [OII] emission line observed in
six AGN hosts at $z\sim0.9$.  These galaxies are a subsample of AGN hosts
detected in the Cl1604 supercluster that exhibit strong Balmer absorption
lines in their spectra and appear to be in a post-starburst or post-quenched
phase, if not for their [OII] emission.  Examining the flux ratio of the
[NII] to H$\alpha$ lines, we find that in five of the six hosts the dominant
source of ionizing flux is AGN continuum emission.  Furthermore, we find that
four of the six galaxies have over twice the [OII] line luminosity that could
be generated by star formation processes alone given their H$\alpha$ line
luminosities.  This strongly suggests that AGN-excited narrow-line emission
is contaminating the [OII] line flux.  A comparison of star formation rates
calculated from extinction-corrected [OII] and H$\alpha$ line luminosities indicates that the
former yields a five-fold overestimate of current activity in these
galaxies.  Our findings reveal the [OII] line to be a poor indicator of star
formation activity in a majority of these moderate-luminosity Seyferts.
This result bolsters our previous findings that an increased fraction of AGN
at high redshifts are hosted by galaxies in a post-starburst phase.  
The relatively high fraction of AGN hosts in the Cl1604
supercluster that show signs of recently truncated star formation activity
suggest AGN feedback may play an increasingly important role in suppressing
ongoing activity in large-scale structures at high redshift.   
\end{abstract}


\section{Introduction}

There is growing evidence for a link between the evolution of supermassive
black holes (SMBHs) and the galaxies that host them.  This includes the observed
correlation between SMBH mass and bulge velocity dispersion (Gebhardt et
al.~2000; Tremaine et al.~2002).
How such correlations are established remains to be determined, but they
suggest that SMBHs and their host galaxies play a role in regulating one
another's growth.  Recent hydrodynamical simulations suggest that feedback
from active galactic nuclei (AGN) may provide a means for this regulation.
In this scenario, a sufficiently energetic AGN can drive outflows that
disrupt its host galaxy and effectively quench ongoing star formation by
removing the galaxy's gas supply (Springel et al.~2005; Hopkins et al.~2005).

If AGNs are responsible for rapidly quenching star formation in their hosts,
then signs of recent transformation should be present in these galaxies,
including post-starburst spectral signatures.  Post-starburst galaxies
exhibit Balmer absorption features due to the presence of recently formed
($<$1 Gyr) A-type stars, but lack emission lines that are indicative of
ongoing star formation, such as H$\alpha$ at 6563\AA\ or the [OII] doublet at
3727\AA.  Similar, yet weaker, features are also found in galaxies that have
experienced a rapid truncation of normal star formation (i.e.~post-quenched
galaxies; Poggianti 2004).  

Several recent studies have found evidence that post-starburst features are
detected in AGN hosts more often than in their non-active counterparts.  For
example, Goto (2006) finds that 4.2\% of optically identified AGNs at low
redshifts exhibit post-starburst spectral signatures, compared to only 0.2\%
of normal galaxies.  At higher redshifts, Georgakakis et al.~(2008)
determined that $21\%$ of X-ray selected AGN hosts located on the red
sequence at $0.68 \le z \le 0.88$ exhibit post-starburst signatures.  At
slightly higher redshifts, Kocevski et al.~(2009a, hereafter Koc09a) report
that $\sim44\%$ of the X-ray selected AGNs found in the Cl1604 supercluster
at $z\sim0.9$ show strong Balmer absorption features. 

\begin{center}
\tabletypesize{\scriptsize}
\begin{deluxetable*}{cccccccc}
\tablewidth{0pt}
\tablecaption{Properties of the Cl1604 AGN Sample Observed with NIRSPEC\label{tab-agn}}
\tablecolumns{8}
\tablehead{\colhead{} & \colhead{} & \colhead{RA} & \colhead{Dec} &
           \colhead{} &  \colhead{$L_{\rm x}$ (0.5-8 keV)} & \colhead{} & \colhead{} \\  
           \colhead{ID} & \colhead{Name} & \colhead{(J2000)} &  \colhead{(J2000)} &  
           \colhead{$z$} & \colhead{($\times10^{43}$)$^{\dagger}$} &
           \colhead{[OII] EW} & \colhead{log(F[NII]/F[H$\alpha$])} } 
\startdata
 X0  & J160415.6+431016  &  241.06492  &  43.17134  &  0.900  &  4.47  & -16.8 & -0.49   \\ 
 X1  & J160425.9+431245  &  241.10783  &  43.21265  &  0.871  &  1.31  & -13.7 & $>$-0.12$^{\ddagger}$   \\ 
 X2  & J160423.9+431125  &  241.09959  &  43.19052  &  0.867  &  3.39  & -7.5  & -0.22   \\ 
 X3  & J160436.7+432141  &  241.15290  &  43.36147  &  0.923  &  2.28  & -7.0  & -0.06   \\ 
 X4  & J160408.2+431736  &  241.03416  &  43.29359  &  0.937  &  2.75  & -7.2  &  0.60   \\  
 X5  & J160401.3+431351  &  241.00546  &  43.23089  &  0.927  &  3.44  & -3.3  & $>$0.55$^{\ddagger}$   \\  
\vspace*{-0.075in}
\enddata
\tablecomments{$^{\dagger}$ in units of $h_{70}^{-2}$ erg s$^{-1}$,
  $^{\ddagger}$ $3\sigma$ lower limit}
\end{deluxetable*}
\end{center}

\vspace{-0.3in}
One of the difficulties of extending such work to higher redshifts is that
the H$\alpha$ line moves out of the optical window, forcing surveys to
rely on the [OII] doublet as the primary indicator of star formation
activity.  However, a comprehensive study by Yan et al.~(2006) suggests that
[OII] emission is a poor indicator of star formation activity in many
galaxies.  They show that in 91\% of red, early-type galaxies that exhibit
[OII] emission, the line originates from AGN activity and not normal star
formation processes.  This poses significant problems for the correct
identification of post-starburst/post-quenched galaxies at high redshift,
which are often selected based on the absence of [OII] emission
(e.g.~Dressler \& Gunn 1983, Poggianti et al.~1999).   If AGNs contribute
significantly to their host galaxies' [OII] line flux, then excluding all
[OII] emitters from such samples would severely underestimate the fraction of
AGN hosts that are truly in a post-starburst or post-quenched phase.  

In this study we use emission line diagnostics measured from optical and near
infrared spectra to investigate the origin of [OII] emission in six X-ray
selected AGNs at $z\sim0.9$. These AGNs are a subset of active galaxies
detected in the Cl1604 supercluster by Koc09a that were
found to have both strong Balmer absorption lines indicative of
post-starburst/post-quenched systems as well as moderate levels of [OII]
emission.  If the observed [OII] lines in these galaxies are of AGN origin,
then the high incidence of Balmer features would indicate an increased
post-starburst fraction among AGN hosts at high redshift. Using newly
obtained Keck II Near-Infrared Echelle Spectrograph (NIRSPEC) spectroscopy,
we examine the flux ratio of the [NII] $\lambda6584$\AA$ $ to H$\alpha$ lines
to ascertain the source of the ionizing flux exciting the [OII] emission found
in these galaxies. 

We present our analysis in the following manner: \S2 describes the Cl1604
supercluster and our AGN sample in greater detail and \S3 discusses the
observation and reduction of the near-infrared spectroscopy.  In \S4 we
describe our methods for measuring equivalent widths and line flux ratios. In
\S5 we present and discuss our findings, while in \S6 we summarize our conclusions.

\section{The Cl1604 AGN Sample}

The six AGNs targeted for NIRSPEC observations are spectroscopically confirmed
members of the Cl1604 supercluster, a large-scale structure which consists of
eight galaxy clusters and groups at a median redshift of $z\sim0.9$ (Gal \& Lubin 2004).   
The supercluster has been extensively studied at a variety of wavelengths
(Gal et al.~2008, Kocevski et al.~2009b, 2010, Lubin et al.~2010) and in 
Koc09a we matched nine X-ray sources to galaxies within the structure. 
The environments, morphologies and spectral
properties of these galaxies are discussed in detail in Koc09a.  The AGNs have
moderate X-ray luminosities ($L_{\rm X} \sim 10^{43}$ erg s$^{-1}$) and their
optical spectra are consistent with type 2 Seyferts.  
The average spectral properties of the Cl1604 AGNs targeted with NIRSPEC can be seen in Figure
\ref{fig-stacked_spec}, where a high-S/N composite spectrum created by
co-adding the individual AGN spectra is shown.   
All of the AGNs exhibit moderate to strong [OII] emission lines and a majority show
strong Balmer absorption features, which are clearly visible in the stacked spectra.

Our available observing time allowed for only two-thirds of the full Cl1604
X-ray AGN sample to be targeted with NIRSPEC.  Preference was given to 
those galaxies whose H$\alpha$ and
[NII] features would be detected blueward of the strong OH airglow lines at
$\lambda\approx1.27\mu$m ($z\leq0.93$).  The details of the NIRSPEC
observations are  given in \S\ref{NIR_spec}. 
The coordinates, redshifts, X-ray luminosities, and spectral properties of
the six Cl1604 AGNs observed with NIRSPEC are listed in Table \ref{tab-agn}.

\section{Observations and Data Reduction}

\subsection{Optical Spectroscopy}
\label{optical_spec}

The vast majority of redshifts in the Cl1604 field come from observations of
18 slitmasks with DEIMOS on the Keck II telescope between May 2003 and
June 2010. The details of the majority of the observations and spectroscopic
selection are described in  Gal et al.~(2008).  Briefly, slitmasks were
observed with the 1200 l mm$^{-1}$ grating with a FWHM resolution of
$\sim$1.7\AA\ (68 km s$^{-1}$) and a typical wavelength coverage of 6385\AA\
to 9015\AA.\  The exposure frames for each slitmask were reduced using the
DEEP2 version of the \emph{spec2d} package (Davis et al.~2003) as described
in Lemaux et al.~(2009). 

In total, 1339 high-quality extragalactic DEIMOS spectra were obtained in the
Cl1604 field, with 432 objects having measured redshifts within the adopted
redshift range of the supercluster.  Combined with additional
redshifts obtained with the Low-Resolution Imaging Spectrometer (Oke et
al. 1995), 517 high quality spectra have been obtained for members of the
Cl1604 supercluster.

\subsection{Near-Infrared Spectroscopy}
\label{NIR_spec}

To measure the H$\alpha$ and [NII] emission lines we used the
NIRSPEC spectrograph (McLean et al. 1998) on the Keck II
telescope. Five of our Cl1604 X-ray AGNs were targeted in June 2009 and one in
June 2007 as part of the sample in Lemaux et al.~(2010, hereafter Lem10). 
Observations were taken in low-resolution mode with slit widths of
0.76{\arcsec}, resulting in a pixel scale of 3 \AA\ pix$^{-1}$ and a FWHM
resolution of $\sim$8\AA. The observations were taken through the NIRSPEC-3
filter, with a typical wavelength coverage of 2900 \AA\ and central
wavelength of 1.273 $\mu$m.  Conditions were photometric and seeing ranged
from 0.55$\arcsec$-0.7$\arcsec$.  

\begin{figure}[t]
\epsscale{1.1}
\plotone{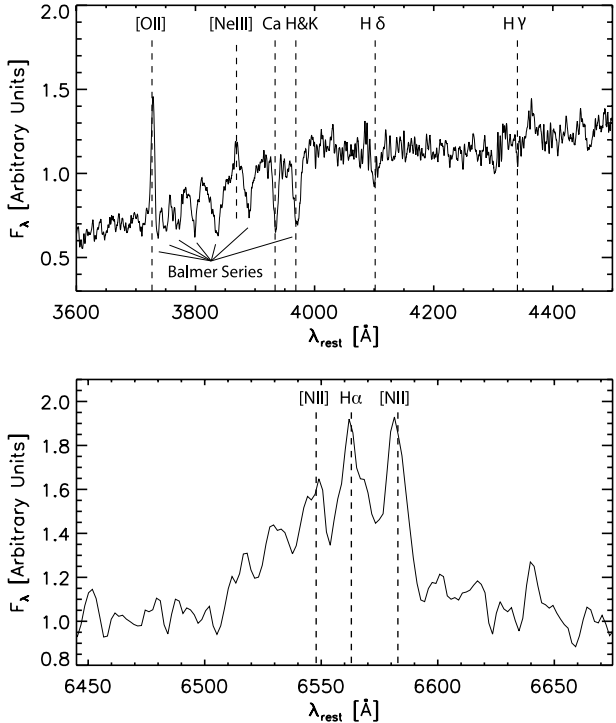}
\caption{\label{fig-stacked_spec} Stacked DEIMOS (\emph{above}) and NIRSPEC
  (\emph{below}) rest-frame composite spectra the Cl1604 AGN sample. }
\end{figure}

The observation of each setup consisted of staggered exposures of 300s to
900s between nods on the sky  of 1.4$\arcsec$ - 2.5$\arcsec$ along the
42$\arcsec$ slit, with total integration times varying between 1800 and 3600s.
Two standard stars drawn from the UKIRT list of bright
standards\footnote{http://www.jach.hawaii.edu/UKIRT/astronomy/calib/phot\_cal/
  ukirt\_stds.html} were observed, HD105601 (A2) at evening twilight and
HD203856 (A0) at morning twilight.  

The NIRSPEC data were reduced using a publicly available semi-automated
Interactive Data Language (IDL) pipeline (George Becker, private
communication). This pipeline is used to generate flat-field,
dark-subtracted, cosmic ray cleaned, wavelength and flux calibrated one- and
two-dimensional spectra. The details of this pipeline as well as
further details on the reduction process are given in Lem10. 
Composite NIRSPEC and DEIMOS spectra of the six Cl1604 AGN hosts, created by
averaging the spectra of the individual galaxies, are shown in the two panels
of Figure \ref{fig-stacked_spec}.

\section{Analysis}

\subsection{Spectral Line Measurements}

In each processed one-dimensional DEIMOS and NIRSPEC spectrum, we measure the
rest-frame EW of the [OII] and H$\alpha$ nebular emission features using two
techniques: bandpass measurements and line-fitting techniques.  
Bandpass measurements were performed on all spectra by integrating around the
vicinity of a spectral feature using identical bandpass endpoints to
those in Lem10.
Line-fitting was performed on all spectra where emission lines were detected
at a significance of greater than 3$\sigma$.  
For each spectrum, the EW measurement was chosen from the better of the two
methods, typically depending on the feature S/N.  The convention adopted in
this paper is for positive EWs to correspond to features observed in emission
and negative EWs to  those observed in absorption.  Emission line fluxes of
the [OII], H$\alpha$ and [NII] features were calculated using the same two
methods as those used for EW measurements.  While we include EW measurements
detected in emission at a significance less than 3$\sigma$ in our analyses,
line fluxes measured at low significance were assigned 3$\sigma$ upper
limits. Errors are calculated either through a combination of the covariance
matrix and Poisson errors (for bandpass measurements) or simply through the
covariance matrix of the fit (for line-fitting techniques). 

Absolute flux calibration and slit throughput corrections for the Cl1604
DEIMOS and NIRSPEC data were obtained in a manner nearly identical to that of
Lemaux et al.~(2009) and Lem10, respectively.  The NIRSPEC calibration
involved comparing observations of the standard star HD203856 with a scaled
flux calibrated spectrum of $\alpha$Lyr (Colina, Bohline, \& Castelli 1996).
Following the analysis of Lem10, an internal extinction of $E(B-V)$ = 0.3 was
adopted for all galaxies in our sample and corrected for using the Calzetti
et al.~(2000) reddening law.

\begin{figure}[t]
\epsscale{1.22}
\plotone{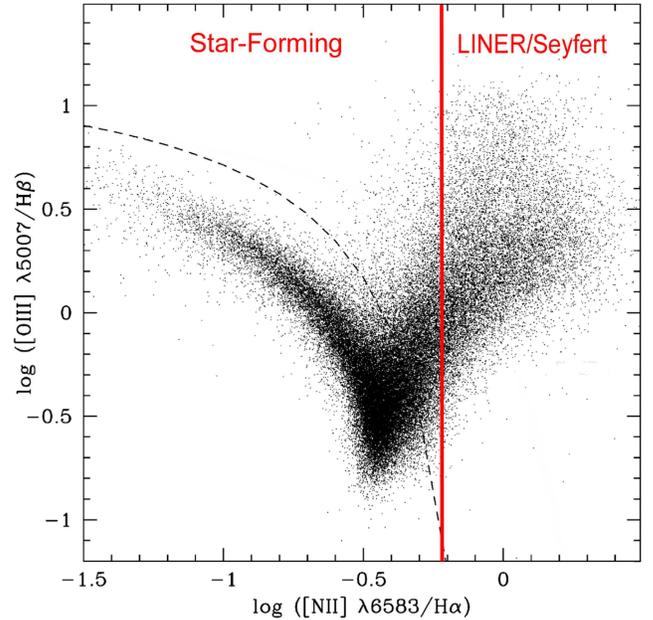}
\caption{\label{fig-SDSS_BPT} Figure adopted from Kauffmann et al.~(2003)
  showing the emission line flux ratios for ∼55,000 SDSS galaxies.
  The dashed line denotes the dividing line between star-forming galaxies and
  those with emission from a LINER or a Seyfert defined by Kauffmann et
  al. The red solid vertical line shows our adopted boundary of
  log(F[NII]/F[H$\alpha$]) = -0.22 between star-forming and LINER/Seyfert
  galaxies.} 
\end{figure}

\subsection{Line Ratio Diagnostics}
\label{Sect-line_ratios}

Several sources of photoionizing radiation can give rise to narrow emission
lines in galaxies, including young stellar populations and AGN-related
continuum emission.  Fortunately, the different line species these two
mechanisms excite provide a means to distinguish which is the dominant source of
ionizing flux.  Therefore it has become common to use the relative strength
of lines with high and low ionization potentials to differentiate these two
excitation mechanisms.

\begin{figure}[t]
\epsscale{1.1}
\plotone{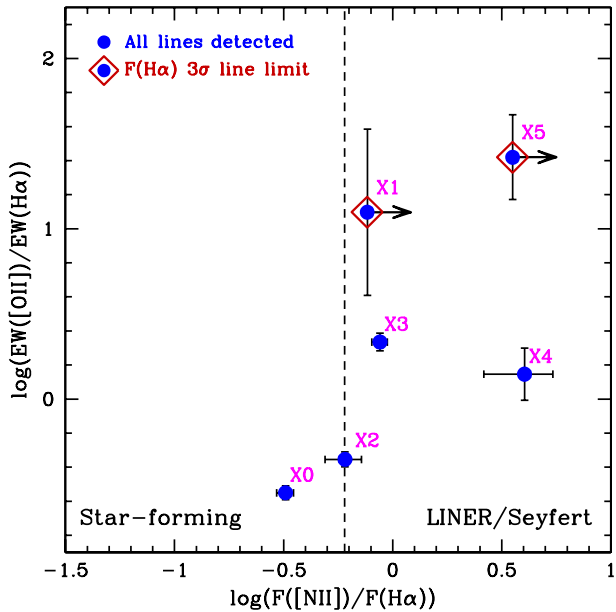}
\caption{\label{fig-Cl1604_BPT1} Ratio of [OII] and H$\alpha$ EWs as a
  function of F([NII])/F(H$\alpha$) for the six Cl1604 AGNs observed with
  NIRSPEC.  Galaxies with $3\sigma$ upper limits on F(H$\alpha$) are plotted
  with arrows.  The vertical dashed line at log(F[NII]/F[H$\alpha$] = -0.22
  indicates our dividing line between star-forming and “LINER/Seyfert”
  galaxies. }
\end{figure}

Introduced by Baldwin, Phillips, \& Terlevich (1981; hereafter BPT), such line ratio diagnostics
typically utilize the strength of the [OIII] feature relative to H$\beta$ and
the strength of a forbidden line (typically 6300\AA\ [OI], [NII], or
6716\AA+6731\AA\ [SII]) relative to H$\alpha$.  An example of a
BPT-diagram using the [NII]/H$\alpha$ and [OIII]/H$\beta$ line flux ratios as
measured in $\sim55,000$ SDSS galaxies is shown in Figure \ref{fig-SDSS_BPT}.  
In this parameter space, Seyferts and “low-ionization nuclear
emission-line regions” (LINER) dominated sources (hereafter LINER/Seyfert)
have higher values of [NII]/H$\alpha$.
In a study of nearly 100,000 SDSS galaxies, Kewley et al.~(2006) found that
the logarithm of the F([NII])/F(H$\alpha$) ratio (hereafter F$_{\rm
  [NII]/H\alpha}$) for star-forming galaxies varied from -1.5 in extremely
metal-poor systems to -0.3 in those with super-solar abundances. 
This upper bound is similar to the maximal boundary 
of log(F$_{\rm [NII]/H\alpha}$) $ = -0.22$ found for star-forming galaxies in a
sample of  Kauffmann et al.~(2003); this limit is denoted by the vertical
line in Figure \ref{fig-SDSS_BPT}.

For this study we lack observations of the H$\beta$ and [OIII] emission
features due to the redshift of our target galaxies.  Instead, we rely on a
pseudo-BPT diagram that utilizes the F$_{\rm
  [NII]/H\alpha}$ ratio as the primary discriminator between star-forming galaxies
and galaxies dominated by LINER/Seyfert emission.  In addition, we use the EW
of the [OII] and H$\alpha$ emission features as a replacement for the
traditional ordinate of the BPT diagram (i.e., F([OIII])/F(H$\beta$)).  
Yan et al.~(2006) previously demonstrated the effectiveness of this type of
pseudo-BPT diagram in separating star-forming and AGN-like systems (they
utilized the [NII]/H$\alpha$ and [OII]/H$\beta$ ratios; see their Figure 11). 

\begin{figure}[t]
\epsscale{1.18}
\plotone{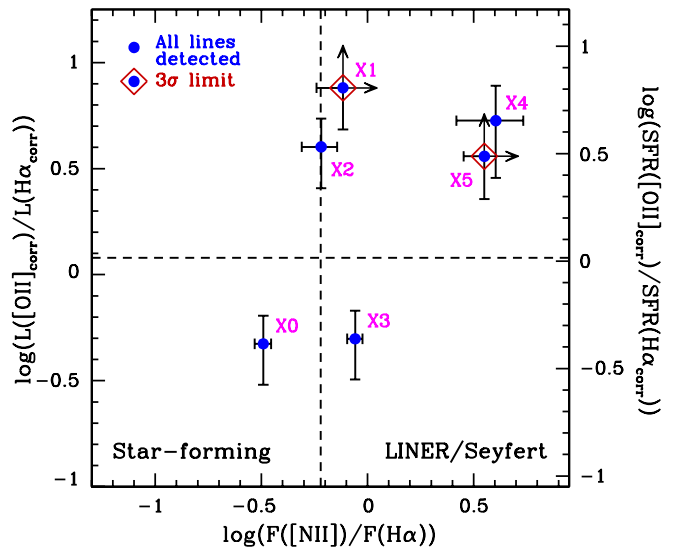}
\caption{\label{fig-Cl1604_BPT2} The ratio of extinction
  corrected L([OII]) and L(H$\alpha$) as a function of
  log(F[NII]/F[H$\alpha$]) for the Cl1604 AGN sample. 
  Galaxies with 3$\sigma$ upper limits on F(H$\alpha$) and L(H$\alpha$) are
  plotted with arrows. The vertical dashed line at F[NII]/F[H$\alpha$] =
  -0.22 denotes our boundary between a star-forming and a LINER/Seyfert
  classification. The horizontal dashed line at log(L[OII]/L[H$\alpha$]) =
  0.08 is the average extinction corrected luminosity ratio for star-forming
  galaxies at low-redshift (Kewley et al. 2004).}
\end{figure}

Without the additional information that F$_{\rm [OIII]/H\beta}$ provides, we
have chosen to divide star-forming galaxies from those dominated by a
LINER/Seyfert component at the Kauffmann et al.~(2003) limit of log(F$_{\rm
  [NII]/H\alpha}$) $ = -0.22$.  This value is similar to cuts used in other
studies when H$\beta$ and [OIII] are weak or unobservable (e.g., Miller et
al.~2003; Stasi{\'n}ska et al.~2006; Lem10).  Furthermore, we
believe this value to be a conservative boundary between star-forming
galaxies and LINER/Seyferts as previous studies have found evidence for
substantial AGN contamination out to as far as log(F$_{\rm [NII]/H\alpha}$)$
= -0.35$ (Stasi{\'n}ska et al.~2006).  In the following section we discuss
where the six Cl1604 X-ray AGNs targeted with NIRSPEC fall on our pseudo-BPT
diagram.

\vspace{0.1in}
\section{Results}

In Figure \ref{fig-Cl1604_BPT1} we plot the ratio of [OII] and H$\alpha$
rest-frame EWs as a function of F$_{\rm [NII]/H\alpha}$ for the six Cl1604
AGN hosts observed with NIRSPEC.  
In four of these six galaxies we detect strong ($>3\sigma$) [OII], [NII], and
H$\alpha$ emission lines, while in two galaxies (X1 and X5) [NII] was
detected in our NIRSPEC spectroscopy, but H$\alpha$ was not.  For these two
galaxies we make use of $3\sigma$ upper limits on the flux and luminosity of
the H$\alpha$ line.  In both cases the $3\sigma$ limits are such that a
classification could be made based on our pseudo-BPT diagram. 
We find that five of the six Cl1604 AGN hosts have [NII]/H$\alpha$ flux
ratios greater than log(F$_{\rm [NII]/H\alpha}$) $ = -0.22$, the maximum
value found for star-forming galaxies by Kauffmann et al.~(2003).  This
suggests the dominant source of ionizing flux in these galaxies is
AGN-related emission and not radiation from young stellar populations. 

The single Cl1604 AGN found to lie in the star-forming region of our pseudo-BPT
diagram is hosted by a spiral galaxy with clear signs of ongoing star
formation within its disk.  This galaxy has the latest morphological type
among the Cl1604 AGN hosts.  An ACS-F814W image of the galaxy can be seen in
Figure 4 of Koc09a (see source 5).  The galaxy also appears to be in the
early stages of an ongoing merger, as it has a nearby companion that is
spectroscopically confirmed to be at the same redshift (Koc09a).  Given these
properties, it is not surprising (and somewhat reassuring) that our emission
line diagnostic indicates the presence of ongoing star formation in the system. 

To further investigate whether the [OII] emission lines observed in these
galaxies could originate from star formation processes, we compared the
extinction-corrected [OII] and H$\alpha$ line luminosities measured in each galaxy.
The L([OII])/L(H$\alpha$) (hereafter L$_{\rm [OII]/H\alpha}$) ratio, plotted
against F$_{\rm [NII]/H\alpha}$, is shown Figure \ref{fig-Cl1604_BPT2}.
Kewley et al.~(2004) previously demonstrated that the maximum L$_{\rm
  [OII]/H\alpha}$ ratio that can be achieved by star-forming regions is roughly 2.1.  
We find that four of the six Cl1604 AGN hosts have L$_{\rm [OII]/H\alpha}$ 
ratios that exceed this limit.  On average, these galaxies have over twice the [OII] line
luminosity that could be generated by star formation processes alone given their
observed H$\alpha$ line luminosity.  The two galaxies that fall below the
Kewley et al.~maximum are sources X0, the star-forming spiral host previously
discussed, and source X3.  This latter source is the only Cl1604 AGN in which
the highly ionized [NeV] emission line is visible.  The strength of the [OII]
line is predicted and observed to decrease in high-ionization AGNs (Ho et
al.~1993), which likely explains the lack of [OII] contamination in this galaxy.

To determine the degree to which current star formation activity would be
overestimated in the galaxies showing excess [OII] emission, we calculated
the star formation rate (SFR) of each host using the relations of Kennicutt
(1998) and the observed [OII] and H$\alpha$ line luminosities.  In Figure
\ref{fig-Cl1604_BPT2} we compare the SFR derived from the
extinction-corrected H$\alpha$ line to that of the extinction-corrected [OII]
line using a constant $E(B-V)=0.3$.  In the four AGN hosts where [OII] is elevated
relative to H$\alpha$, the SFR would, on average, be overestimated by a factor of five.  
Even if we make no correction for extinction, which attenuates [OII] more
than H$\alpha$, the [OII]-derived SFRs in these galaxies would be over twice
the rate derived from H$\alpha$.

Our finding that the [OII] line can be significantly
contaminated bolster the results of Koc09a, who reported that a large
fraction of the AGNs in the Cl1604 supercluster appear to be associated with
recently quenched hosts.  A major caveat in that study was the assumption
that the observed [OII] emission in the spectra of these galaxies was due to
AGN activity and not ongoing star formation.  We have now demonstrated conclusively
that this assumption is supported by our new observations. This finding
provides significant insight into the nature of these systems.  
In the absence of substantial star formation, the properties of these
galaxies generally agree with AGN feedback scenarios in which AGN-driven
outflows rapidly suppress the star formation activity of their hosts. 
For example, the high incidence of Balmer lines in the spectra of the Cl1604
hosts implies that the last major episode of star formation in many of these
galaxies was truncated within the last $\sim1$ Gyr.  Furthermore, these
galaxies have optical colors that place them between the red sequence of
early-type galaxies and the blue cloud of star-forming objects, while their
morphologies show tell-tale signs of interactions in the recent past,
including fading tidal features and multiple nuclei. 
In most regards, these galaxies resemble a transition population within the
Cl1604 structure that have been caught in the process of transforming 
from actively star-forming galaxies to passively evolving systems.
The fact that a large fraction of AGNs within the Cl1604 supercluster show such
signatures, coupled with recent findings that high-redshift AGNs are more common in
denser environments than they are in the field (Gilli et al 2003; Cappelluti
et al 2005; Eastman et al. 2007; Kocevski et al.~2009b), suggests that
AGN-related feedback may play an important role in accelerating galaxy
evolution in large-scale structures at high redshift.

\section{Conclusions}

We have used emission-line ratios measured from DEIMOS and NIRSPEC  
spectroscopy to determine the origin of the [OII] emission line observed in
six AGN hosts at $z\sim0.9$.  These six galaxies are a subsample of nine AGNs
detected in the Cl1604 supercluster that exhibit strong Balmer absorption
lines in their spectra.  If not for the presence of the [OII] line, many of
these galaxies would be classified as post-starburst or post-quenched systems.

Examining the flux ratio of the [NII] to H$\alpha$ lines, we find that in five
of the six AGN hosts the dominant source of ionizing flux is AGN continuum
emission and not radiation from young stellar populations.  
Furthermore, we find that four of the six galaxies have, on average, over twice the
[OII] line luminosity that could be generated by star formation processes
alone given their H$\alpha$ line luminosities.  This strongly suggests that AGN excited
narrow-line emission is contaminating the [OII] line flux. 
A comparison of SFRs calculated from the [OII] and H$\alpha$ line luminosities
indicates that the former yields a five-fold overestimate of the star
formation activity of these galaxies. 

Our findings reveal the [OII] line to be a poor indicator of current star
formation activity in a majority of these moderate luminosity Seyferts; a
result that agrees with the conclusions of Yan et al.~(2006) and Lem10.  
This has important ramifications on the study of post-starburst and recently
quenched AGN hosts at high redshifts.   Since post-starburst systems are
selected based on the absence of current star formation, any studies that
rely solely on the [OII] line as a measure of current activity in AGN hosts
will severely underestimate the fraction of AGNs found in post-starburst
galaxies. Furthermore, our results confirm the increased connection between AGN
activity and post-starburst/post-quenched systems at high redshifts reported
by Koc09a.  The relatively high fraction of AGN hosts in the Cl1604
supercluster that show signs of recently truncated star formation
activity suggest AGN feedback may play an increasingly important role
in suppressing ongoing activity and accelerating galaxy evolution in
overdense environments at high redshift.


\begin{thebibliography}{}

\bibitem[{Baldwin}, {Phillips}, \&  {Terlevich} 1981]{baldwin81}
{Baldwin}, J.~A., {Phillips}, M.~M., \& {Terlevich}, R. 1981, \pasp, 93, 5

\bibitem[{Calzetti}, {Armus}, {Bohlin}, {Kinney},  {Koornneef}, \& {Storchi-Bergmann} 2000]{calzetti00}
{Calzetti}, D., {Armus}, L., {Bohlin}, R.~C., {Kinney}, A.~L., {Koornneef}, J.,  \& {Storchi-Bergmann}, T. 2000, \apj, 533, 682

\bibitem[{Cappelluti}, {Cappi}, {Dadina},  {Malaguti}, {Branchesi}, {D'Elia}, \& {Palumbo} 2005]{cappelluti05}
{Cappelluti}, N., {Cappi}, M., {Dadina}, M., {Malaguti}, G., {Branchesi}, M.,  {D'Elia}, V., \& {Palumbo}, G.~G.~C. 2005, \aap, 430, 39

\bibitem[{Colina}, {Bohlin}, \& {Castelli} 1996]{colina96}
{Colina}, L., {Bohlin}, R.~C., \& {Castelli}, F. 1996, \aj, 112, 307

\bibitem[{Davis}, {Faber}, {Newman}, {Phillips},  {Ellis}, {Steidel}, {Conselice}, {Coil}, {Finkbeiner}, {Koo}, {Guhathakurta},  {Weiner}, {Schiavon}, {Willmer}, {Kaiser}, {Luppino}, {Wirth}, {Connolly},  {Eisenhardt}, {Cooper}, \& {Gerke} 2003]{davis03}
{Davis}, M., {Faber}, S.~M., {Newman}, J., {Phillips}, A.~C., {Ellis}, R.~S.,  {Steidel}, C.~C., {Conselice}, C., {Coil}, A.~L., {et al.} 2003, in Society of Photo-Optical  Instrumentation Engineers (SPIE) Conference Series, Vol. 4834, Society of  Photo-Optical Instrumentation Engineers (SPIE) Conference Series, ed.  {P.~Guhathakurta}, 161--172

\bibitem[{Dressler} \& {Gunn} 1983]{dressler_gunn83}
{Dressler}, A. \& {Gunn}, J.~E. 1983, \apj, 270, 7

\bibitem[{Eastman}, {Martini}, {Sivakoff}, {Kelson},  {Mulchaey}, \& {Tran} 2007]{eastman07}
{Eastman}, J., {Martini}, P., {Sivakoff}, G., {Kelson}, D.~D., {Mulchaey},  J.~S., \& {Tran}, K.-V. 2007, \apjl, 664, L9

\bibitem[{Gal}, {Lemaux}, {Lubin}, {Kocevksi}, \&  {Squires} 2008]{gal08}
{Gal}, R.~R., {Lemaux}, B.~C., {Lubin}, L.~M., {Kocevksi}, D., \& {Squires},  G.~K. 2008, ArXiv e-prints, 803

\bibitem[{Gal} \& {Lubin} 2004]{gal04}
{Gal}, R.~R. \& {Lubin}, L.~M. 2004, \apjl, 607, L1

\bibitem[{Gebhardt}, {Bender}, {Bower}, {Dressler},  {Faber}, {Filippenko}, {Green}, {Grillmair}, {Ho}, {Kormendy}, {Lauer},  {Magorrian}, {Pinkney}, {Richstone}, \& {Tremaine} 2000]{gebhardt00}
{Gebhardt}, K., {Bender}, R., {Bower}, G., {Dressler}, A., {Faber}, S.~M.,  {Filippenko}, A.~V., {Green}, R., {Grillmair}, C., {et al.} 2000, \apjl, 539, L13

\bibitem[{Georgakakis}, {Nandra}, {Yan},  {Willner}, {Lotz}, {Pierce}, {Cooper}, {Laird}, {Koo}, {Barmby}, {Newman},  {Primack}, \& {Coil} 2008]{georgakakis08}
{Georgakakis}, A., {Nandra}, K., {Yan}, R., {Willner}, S.~P., {Lotz}, J.~M.,  {Pierce}, C.~M., {Cooper}, M.~C., {Laird}, E.~S., {et al.} 2008, \mnras, 385, 2049

\bibitem[{Gilli}, {Cimatti}, {Daddi}, {Hasinger},  {Rosati}, {Szokoly}, {Tozzi}, {Bergeron}, {Borgani}, {Giacconi}, {Kewley},  {Mainieri}, {Mignoli}, {Nonino}, {Norman}, {Wang}, {Zamorani}, {Zheng}, \&  {Zirm} 2003]{gilli03}
{Gilli}, R., {Cimatti}, A., {Daddi}, E., {Hasinger}, G., {Rosati}, P.,  {Szokoly}, G., {Tozzi}, P., {Bergeron}, J., {et al.} 2003, \apj, 592, 721

\bibitem[{Goto} 2006]{goto06}
{Goto}, T. 2006, \mnras, 369, 1765

\bibitem[{Ho}, {Shields}, {Filippenko} 1993]{ho93}
{Ho}, L.~C., {Shields}, J.~C. \& {Filippenko}, A.~V. 2005, \apj, 410, 567

\bibitem[{Hopkins}, {Hernquist}, {Martini}, {Cox},  {Robertson}, {Di Matteo}, \& {Springel} 2005]{hopkins05}
{Hopkins}, P.~F., {Hernquist}, L., {Martini}, P., {Cox}, T.~J., {Robertson},  B., {Di Matteo}, T., \& {Springel}, V. 2005, \apjl, 625, L71

\bibitem[{Kauffmann}, {Heckman}, {Tremonti},  {Brinchmann}, {Charlot}, {White}, {Ridgway}, {Brinkmann}, {Fukugita}, {Hall},  {Ivezi{\'c}}, {Richards}, \& {Schneider} 2003]{kauffmann03}
{Kauffmann}, G., {Heckman}, T.~M., {Tremonti}, C., {Brinchmann}, J., {Charlot},  S., {White}, S.~D.~M., {Ridgway}, S.~E., {Brinkmann}, J., {et al.} 2003, \mnras, 346, 1055

\bibitem[{Kennicutt} 1998]{kennicutt98}
{Kennicutt}, Jr., R.~C. 1998, \araa, 36, 189

\bibitem[{Kewley}, {Geller}, \& {Jansen} 2004]{kewley04}
{Kewley}, L.~J., {Geller}, M.~J., \& {Jansen}, R.~A. 2004, \aj, 127, 2002

\bibitem[{Kewley}, {Groves}, {Kauffmann}, \&  {Heckman} 2006]{kewley06}
{Kewley}, L.~J., {Groves}, B., {Kauffmann}, G., \& {Heckman}, T. 2006, \mnras,  372, 961

\bibitem[{Kocevski}, {Lemaux}, {Lubin}, {Gal},  {McGrath}, {Fassnacht}, {Squires}, {Surace}, \& {Lacy} 2010]{kocevski10}
{Kocevski}, D.~D., {Lemaux}, B.~C., {Lubin}, L.~M., {Gal}, R.~R., {McGrath},  E.~J., {Fassnacht}, C.~D., {Squires}, G.~K., {Surace}, J.~A., {et al.} 2010, ArXiv e-prints

\bibitem[{Kocevski}, {Lubin}, {Lemaux},  {Gal}, {Fassnacht}, {Lin}, \& {Squires} 2009b]{kocevski09a}
{Kocevski}, D.~D., {Lubin}, L.~M., {Lemaux}, B.~C., {Gal}, R.~R., {Fassnacht},  C.~D., {Lin}, R., \& {Squires}, G.~K. 2009a, \apj, 700, 901

\bibitem[{Kocevski}, {Lubin}, {Gal},  {Lemaux}, {Fassnacht}, \& {Squires} 2009a]{kocevski09b}
{Kocevski}, D.~D., {Lubin}, L.~M., {Gal}, R., {Lemaux}, B.~C., {Fassnacht},  C.~D., \& {Squires}, G.~K. 2009b, \apj, 690, 295

\bibitem[{Lemaux}, {Lubin}, {Sawicki}, {Martin},  {Lagattuta}, {Gal}, {Kocevski}, {Fassnacht}, \& {Squires} 2009]{lemaux09}
{Lemaux}, B.~C., {Lubin}, L.~M., {Sawicki}, M., {Martin}, C., {Lagattuta},  D.~J., {Gal}, R.~R., {Kocevski}, D., {Fassnacht}, C.~D., {et al.} 2009, \apj, 700, 20

\bibitem[{Lemaux}, {Lubin}, {Shapley}, {Kocevski},  {Gal}, \& {Squires} 2010]{lemaux10}
{Lemaux}, B.~C., {Lubin}, L.~M., {Shapley}, A., {Kocevski}, D., {Gal}, R.~R.,  \& {Squires}, G.~K. 2010, \apj, 716, 970

\bibitem[{Miller}, {Nichol}, {G{\'o}mez}, {Hopkins}, \&  {Bernardi} 2003]{Miller03}
{Miller}, C.~J., {Nichol}, R.~C., {G{\'o}mez}, P.~L., {Hopkins}, A.~M., \&  {Bernardi}, M. 2003, \apj, 597, 142

\bibitem[{Oke}, {Cohen}, {Carr}, {Cromer}, {Dingizian},  {Harris}, {Labrecque}, {Lucinio}, {Schaal}, {Epps}, \& {Miller} 1995]{oke95}
{Oke}, J.~B., {Cohen}, J.~G., {Carr}, M., {Cromer}, J., {Dingizian}, A.,  {Harris}, F.~H., {Labrecque}, S., {Lucinio}, R., {et al.} 1995, \pasp, 107, 375

\bibitem[{Poggianti}, {Bridges}, {Komiyama}, {Yagi},  {Carter}, {Mobasher}, {Okamura}, \& {Kashikawa} 2004]{poggianti04}
{Poggianti}, B.~M., {Bridges}, T.~J., {Komiyama}, Y., {Yagi}, M., {Carter}, D.,  {Mobasher}, B., {Okamura}, S., \& {Kashikawa}, N. 2004, \apj, 601, 197

\bibitem[{Poggianti}, {Smail}, {Dressler}, {Couch},  {Barger}, {Butcher}, {Ellis}, \& {Oemler} 1999]{poggianti99}
{Poggianti}, B.~M., {Smail}, I., {Dressler}, A., {Couch}, W.~J., {Barger},  A.~J., {Butcher}, H., {Ellis}, R.~S., \& {Oemler}, A.~J. 1999, \apj, 518, 576

\bibitem[{Springel}, {Di Matteo}, \&  {Hernquist} 2005]{springel05}
{Springel}, V., {Di Matteo}, T., \& {Hernquist}, L. 2005, \apjl, 620, L79

\bibitem[{Stasi{\'n}ska}, {Cid Fernandes},  {Mateus}, {Sodr{\'e}}, \& {Asari} 2006]{Stasiska06}
{Stasi{\'n}ska}, G., {Cid Fernandes}, R., {Mateus}, A., {Sodr{\'e}}, L., \&  {Asari}, N.~V. 2006, \mnras, 371, 972

\bibitem[{Tremaine}, {Gebhardt}, {Bender}, {Bower},  {Dressler}, {Faber}, {Filippenko}, {Green}, {Grillmair}, {Ho}, {Kormendy},  {Lauer}, {Magorrian}, {Pinkney}, \& {Richstone} 2002]{tremain02}
{Tremaine}, S., {Gebhardt}, K., {Bender}, R., {Bower}, G., {Dressler}, A.,  {Faber}, S.~M., {Filippenko}, A.~V., {Green}, R., {et al.} 2002, \apj, 574, 740

\bibitem[{Yan}, {Newman}, {Faber}, {Konidaris}, {Koo}, \&  {Davis} 2006]{yan06}
{Yan}, R., {Newman}, J.~A., {Faber}, S.~M., {Konidaris}, N., {Koo}, D., \&  {Davis}, M. 2006, \apj, 648, 281

\end{thebibliography}
\end{document}